# Compensating for charge sharing by a deep-learning method: a preliminary experimental study

Shengzi Zhao, Le Shen, Yuxiang Xing
Department of Engineering Physics, Tsinghua University, China
Key Laboratory of Particle & Radiation Imaging (Tsinghua University), Ministry of Education,
Beijing, China.

*Abstract*—Photon counting detectors (PCDs) bring valuable advantages to diagnostic computed tomography (CT), including lower noise and higher resolution than energy integrating detectors. However, there are still several nonideal factors preventing PCDs from meeting people's expectations, for example, charge sharing and pile up. In this paper, we did some preliminary work on charge sharing and conducted an experimental study using an XCounter PCD to compare the effects of no anti-coincidence, anti-coincidence by hardware and charge sharing compensation by a deep learning method. In our results, a smaller bias and standard deviation are obtained from deep learning method than directly from no-anti-coincidence mode of the detector. Our network also outperforms the anti-coincidence mode of the detector in the low energy bin and has smaller standard deviation in the high energy bin. The results validate that a deep learning method is suitable to compensate for charge sharing.

## I. Introduction

PHOTON counting detectors (PCD) have great potential to improve the quality of spectral computed tomography (CT) images. Smaller pixels in PCD provide the possibility for higher resolution images than EID. A lower threshold in PCD can release the impact of noise from signals with very low amplitudes. Nevertheless, some nonideal factors, including pulse pile-up and charge sharing, can diminish the quality of CT images. Here, we only focus on the charge sharing problem.

There are various software- or hardware-based methods to reduce the impact of charge sharing. For the software-based methods, different models are proposed to fit the energy response of PCDs [1], [2]. These models calibrate model parameters to obtain accurate spectral distortion information. The hardware-based methods mainly include Charge Sharing Addition (CSA) methods and Charge Sharing Discrimination (CSD) methods [3], [4], [5]. The key idea of these two methods is to add circuits in a PCD to identify and process the signals of charge sharing events. However, CSA methods decrease the counting rate of PCD, and CSD methods decrease its detection efficiency. Recently, a new type of detectors with extra coincidence counters is proposed [6], [7]. The idea of these detectors is to record the counts of charge sharing events and process the data after detection, which will not reduce detection efficiency. Deep learning has been a successful method in many fields and can also be used to train a network on large datasets to recognize patterns indicative of charge sharing [8]. There are also published works using deep learning to correct systematic errors caused by non-ideal factors including charge sharing [9], [10]. After neural networks trained, they can then be used to correct for charge sharing and spectrum distortion in new data.

In this work, we study the effect of a deep learning method for charge sharing compensation by a real experiment on an XCounter PCD. In the following, we firstly introduce our method briefly. Then, we introduce the configurations of our experiments and present the results afterwards. Discussion and Conclusion are given in the last section.

## II. Method

The XCounter detector has two modes. The anti-charge-sharing mode (Anti) will activate the circuits to compensate for charge sharing with the CSA method. The no anti-charge-sharing mode (noAnti) will record the counts without any process. We denote the detected photon counts with and without objects on a ray path as $N_c^{\text{det}}$ and $N_c^{\text{det,air}}$ with the subscript $c \in \{L, H\}$ representing two energy bins. Both noAnti data and Anti data can be processed by:

$$Y_{c,i}^{\text{det}} = -\ln \frac{N_{c,i}^{\text{det}}}{N_{c,i}^{\text{det,air}}}, \ c \in \{L, H\}. \qquad (1)$$

to obtain the attenuation integrals at corresponding effective energies of the two energy bins.

Assuming uniform responses among detector pixels, we utilize a neural network to compensate for charge sharing pixel by pixel. Considering a pixel of interest (POI), our network processes the counts of a small detector patch with the size of $n_p \times n_p$ centered on the POI and outputs effective attenuation lengths in the two energy bins corresponding to the POI. For a conventional PCD with two energy bins, the size of inputs is $2 \times n_p \times n_p$. A CNN type network architecture similar with the network in [10] is used.

We conducted an experiment on an imaging system with an XCounter PCD in our laboratory, as shown in Fig. 1. PMMA

blocks combined with an aluminum block of different thicknesses, as shown in Fig. 1(b), were scanned to get a training dataset. The thicknesses of the PMMA blocks are from 6 mm to 52 mm with a step of 2 mm, and the thicknesses of the aluminum block are from 2 mm to 20 mm with a step of 2 mm. A PMMA cylinder with 12 aluminum cylindrical inserts, as shown in Fig. 1(c), was scanned to get data independent of the training data for testing. The diameter of the PMMA cylinder is 50 mm, and of the aluminum cylinder is 5 mm. For data acquisition, the X-ray source is set to 130 kVp and 0.3 mA, the source-to-detector distance is 455 mm, and the source-to-isocenter distance is 316 mm. As for energy bins, we choose a narrow low energy bin [40, 50] keV to avoid significant beam hardening when calculating effective attenuation lengths. We choose [50, 130] keV for the high energy bin because beam hardening is not significant in the high energy bin, and the right edge of the interval is the X-ray source's voltage. Each detector pixel is 0.1mm × 0.1mm, and there are 128 × 1536 pixels in total. We scanned the two phantoms at 360 views uniformly distributed in $2\pi$. For each view, we acquire ten frames in a second. We randomly choose patches of size $5 \times 5$ on all projections, i.e., $n_p = 5$ in our real experiments. In total, 1,564,800 patches and 190,000 patches in training and test datasets are extracted, respectively.

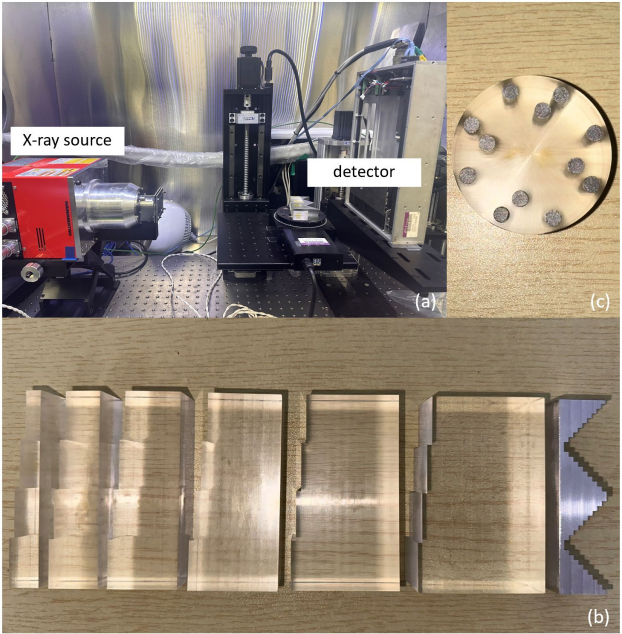

Fig. 1. (a) Photograph of the benchtop photon-counting CT imaging system. (b) Photograph of the phantoms for training data acquisition. (c) Photograph of the phantoms for the test data acquisition.

In our experiment, only the noAnti data is used as input to train our network. We calculate effective attenuation lengths as Eq. (2) and use them as labels for training the network and calculating the metrics in Eqs. (3) and (4).

$$Y^*_{c,\text{POI}} = -\ln \frac{\int_{\text{energy bin } c} S_{\text{eff}}(E) \exp\left(-\int_{l_{\text{POI}}} \mu(E,l) \mathrm{d}l\right) \mathrm{d}E}{\int_{\text{energy bin } c} S_{\text{eff}}(E) \mathrm{d}E} \quad (2)$$

Here, $S_{\text{eff}}(E)$ is the effective spectrum of the X-ray source which we obtained through Monte Carlo simulation based on the voltage, target material, and other configurations in the X-ray source. $\mu(E, l)$ is the attenuation coefficient of the phantom at a position located on a ray path. $\int_{l_{\text{POI}}} \mu(E,l) \mathrm{d}l$ is the integration of the attenuation coefficients along the ray path $l_{\text{POI}}$ corresponding to a POI.

Two quantitative metrics, the average of absolute normalized bias (AANBias) and the average of normalized standard deviation (ANSTD) are utilized to evaluate the results,

$$\text{AANBias} = \frac{1}{2N} \sum_{n=1}^{N} \sum_{c \in \{L,H\}} \frac{\left|\frac{1}{N_{\text{sample}}} \sum_{s=1}^{N_{\text{sample}}} Y_{c,n,s} - Y^*_{c,n}\right|}{Y^*_{c,n} + \epsilon_c}, \quad (3)$$

$$\text{ANSTD} = \frac{1}{2N} \sum_{n=1}^{N} \sum_{c \in \{L,H\}} \frac{\sigma(\{Y_{c,n,s}\})}{Y^*_{c,n} + \epsilon_c}. \quad (4)$$

Here, $Y^*_{c,n}$ are the labels of effective attenuation lengths refer to (2) for the $n^{\text{th}}$ PMMA and aluminum thickness, $Y_{c,n,s}$ are the outputs of our method for the $s^{\text{th}}$ noise realization. $N$ is the number of all the different thicknesses, and $N_{\text{sample}}$ is the number of noise realizations for each thickness in the test dataset. $\sigma(\{\bullet\})$ represents the standard deviation of the set $\{\bullet\}$.

Both of the above indicators are relative values. When the thicknesses are very small, especially when approaching zero, a small absolute error could lead to a significant relative error, which is not conducive to measuring the performance of our method. Therefore, we ignored samples with effective attenuation lengths close to zero for now.

### III. Experimental Results

Results of AANBias and ANSTD (error bar) are plotted in Fig. 2. We firstly focus on the results obtained by directly processing the detected data refer to (1). For noAnti data, the effective attenuation length $Y_c^{\text{det}}$ directly computed from $N_c^{\text{det}}$ is of big error in both energy bins because of charge sharing. Higher bias in the low energy bin implies it is more severely affected by charge sharing than the high energy bin. After hardware correction, the results obtained from the Anti data show a significant reduction in bias. In addition, hardware correction mitigates the impacts that charge sharing reduces the counts in the high energy bin and increases the counts in the low energy bin. Therefore, there are fewer counts in the low energy bin for Anti data than those for noAnti data, leading to a higher standard deviation for Anti data than that for noAnti data. The standard deviation in the high energy bin could also be explained similarly.

By utilizing the network to process noAnti data, both the bias and standard deviation are significantly reduced in the two energy bins compared with the results directly from noAnti data. That means our method has the capability to compensate for charge sharing. In the low energy bins, our network achieves better results than that directly from Anti data. The reason might be that there are multiple nonideal factors in experiments, such as nonuniform detector responses among different pixels and beam hardening that we are not able to isolate in real experiment. Our network could also compensate for those problems to some extent because it does compensation comprehensively. In the high energy bin, the hardware compensation has a smaller bias than our method but has a larger standard deviation.

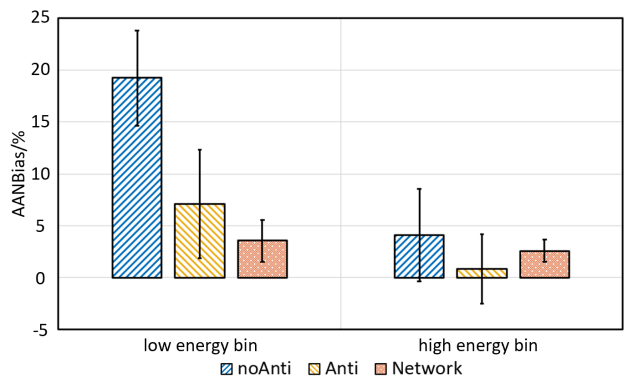

Fig. 2. Results from experiments. The error bars mean ANSTD. "Anti" and "noAnti" mean the results are obtained directly referring to (3). "Network" means the results are estimated using our deep learning method.

## IV. DISCUSSION

Our method processes the counts of a patch centered on a POI and outputs effective attenuation integrals corresponding to the POI. The corrected counts are from comprehensively taking use of information in a patch. The results from Anti data is directly computed from data of POI that is with the circuits utilizing the counts of neighboring pixels of POI to correct charge sharing.

In this practical experiment, there are many nonideal factors that are difficult to isolate and has to be considered in future. Firstly, the spectrum in Eq. (2) is from MC simulation. Although we configured MC simulation according to the real parameters of X-ray source, it is not possible that the simulated spectrum perfectly matches the real spectrum exactly. Secondly, The detection efficiency of the detector varies with the energy of X-ray photons and differs between pixels. It's not easy to get it measured and accounted. Thirdly, pulse pile-up events are mixed with charge sharing events. These factors pose significant difficulties to us to isolate and consider the impact of charge sharing only. Therefore, the preliminary results can only shed some light on the possible solution for charge sharing compensation. Our future work is to further study deep learning method for non-ideal factor compensation in PCD and validate our network more comprehensively.

## V. CONCLUSION

In this work, we conduct a preliminary experimental study to test our method on a conventional PCD with two energy bins. The results demonstrate the effectiveness of our deep learning method for charge sharing compensation. It also provides a potentially feasible way to construct training datasets in practical situation to meet deep learning requirement.

ACKNOWLEDGMENT

This work is funded by National Natural Science Foundation of China under NNSFC 12275151. This work is also supported by Center of High Performance Computing, Tsinghua University.